# Large low-field positive magnetoresistance in nonmagnetic half-Heusler ScPtBi single crystal

Zhipeng Hou, Yue Wang, Enke Liu, Hongwei Zhang, Wenhong Wang [a)] and Guangheng Wu

State Key Laboratory for Magnetism, Beijing National Laboratory for Condensed Matter Physics, Institute of Physics, Chinese Academy of Sciences, Beijing 100190, China

**Abstract**

High-quality nonmagnetic half-Heusler ScPtBi single crystals were synthesized by a Bi self-flux method. This compound was revealed to be a hole-dominated semimetal with a large low-field magnetoresistance up to 240% at 2K in a magnetic field of 1T. Magneto-transport measurements demonstrated that the large low-field magnetoresistance effect resulted from the coexistence of field-induced metal-semiconductor transition and weak-antilocalization effect. Moreover, Hall measurements indicated that ScPtBi single crystal showed a high mobility over a wide temperature region even up to room temperature (4050 $cm^2V^{-1}s^{-1}$ at 2K - 2016 $cm^2V^{-1}s^{-1}$ at 300K). These findings not only suggest the nonmagnetic ScPtBi semimetal a potential material candidate for applications in high-sensitivity magnetic sensors, but also are of great significance to comprehensively understand the rare-earth based half-Heusler compounds.

Correspondence and requests for materials should be addressed to W.H.W (email: wenhong.wang@iphy.ac.cn)

Recently, searching for new materials with semimetallic features in electronic structure protected by the interplay of symmetry and topology has attracted increasing attention, following the observation of large non-saturation magnetoresistance (MR) in a variety of nonmagnetic semimetals such as $Cd_3As_2$,[1] TaAs,[2] NbP,[3] and $WTe_2$.[4] Electronic transport in heavy half-Heusler compounds is distinguished from others by the fact that the strong spin-orbit coupling (SOC) induces many fascinating phenomena, such as the heavy fermion behavior,[5] noncentersymetry superconductors,[6-8] possible topological insulators,[9-13] large positive MR,[14-16] and extremely high mobility,[17] which are of great significance to the applied physics field.

According to band structure calculations, half-Heusler ScPtBi is predicted to show similar band inversion to that of HgTe.[9, 10, 13] Its topological properties can be created by applying strain or designing an appropriate quantum well structure.[9] Moreover, this material possesses nearly linear dispersion relation of bulk band along with zero gap,[13] which is usually associated with large MR effect and high mobility behavior.[1-4,14, 15, 17, 18] Although the growth of many heavy half-Heusler compounds (such as LuPdBi,[6, 16, 19] LuPtBi,[7] LuPtSb,[15, 20, 21] and YPtBi[8]) have been previously reported, the synthesis of ScPtBi (no matter polycrystal or single crystal) is unsuccessful thus far since this compound is proposed to show a weaker thermodynamic stability.[22, 23] In this work, we report on the growth and magneto-transport properties of high-quality nonmagnetic ScPtBi single crystal. Intriguingly, the metal-semiconductor transition coexists with the weak antilocalization (WAL) effect in this material, which results in a large low-field positive MR (240% at 2K -150% at 100K) in a magnetic field of 1T.

Single crystals of ScPtBi were synthesized by using excess Bi as the flux in a molar ratio of Sc: Pt: Bi = 1:1:10. The starting materials were mixed together and placed in an alumina crucible with the one having higher melting temperatures on the bottom. This process was handled in a glove filled with argon gas. To avoid possible oxidization of the materials and influence of the volatilization of Bi at high temperatures, the whole assembly was firstly sealed inside a Tantalum tube under proper Ar pressure. The Tantalum tube was then sealed into a quartz tube filled with 2 mbar Ar pressure. The crystal growth was carried out in a furnace by heating the tube from room temperature up to 1150℃ over a period of 15h, holding at this temperature for 72 hours, and subsequently slowly cooling to 650℃ at a rate of 1.5℃/h. The excess Bi flux was removed by spinning the tube in a centrifuge at 650℃. After the centrifugation process, most of the flux contamination was removed from the surfaces of crystals and the remaining topical flux was etched by diluted hydrochloric acid.

The as-grown ScPtBi crystals are multi-faceted [Figure 1 (a)] with a preferred orientation along the [111] direction as confirmed by the Laue diffraction in Figure 1 (b). To check the crystal structure, powder x-ray diffraction (PXRD) measurement was carried out on the crushed ScPtBi single crystals using Rigaku x-ray diffractometer with Cu-$K\alpha$ radiation [see Figure 1 (c)]. They crystallize in the MgAgAs-type crystal structure (*F-43m* space group) with a refined lattice parameter $a = 6.50$Å, which is close to the value obtained from the band-structure calculations.[9, 13] The chemical composition was examined by scanning electron microscope with an energy-dispersive x-ray spectrometry (SEM-EDX). This measurement was performed at different positions of a polished crystal surface, and the average chemical composition ratio of Sc: Pt: Bi was established to be 32.7% : 33.0% : 34.3% which is very close to the ideal percentage,[24] confirming the absence of Bi impurities. The transport properties were investigated on a Quantum Design physical properties measurement system (PPMS) by a standard four-terminal method. The longitudinal MR was measured with the magnetic field perpendicular to the current direction. The Hall measurement was carried out by both temperature-swapping and field-swapping modes to ensure the accuracy of our experimental data. All the samples selected for magneto-transport measurements were polished to approximately 0.10 mm in thickness and the current and the magnetic field corresponded to the in-plane and out-of-plane directions of the (111) plane.

Figure 2 (a) shows the temperature dependence of resistivity $\rho_{xx}$ of ScPtBi single crystal under different magnetic fields. In zero field, $\rho_{xx}$ decreases in a metallic manner with the decrease of temperature, and a low residual resistivity of $\rho_{xx}$ (2K) = 0.082 mΩcm is obtained. This value is in the same order with that of YbPtBi,[25] but much lower than that of most reported half-Heusler compounds, such as YPtBi,[8] LuPdBi,[6] LuPtSb,[15, 20] YNiBi,[26] and YPtSb,[27] implying a high sample quality. With the further decrease of temperature, ScPtBi single crystal shows no superconducting transition even at temperature down to 0.3K, which further confirms that there are no nano-scale bismuth inclusions trapped inside crystals.[28] After we apply an external magnetic field, the temperature-dependent behavior of resistivity changes significantly, especially at low temperature. In a low magnetic field of 1T, $\rho_{xx}$ preserves the metallic behavior but the slope decreases with the decrease of temperature from 300K. At about 260K, $\rho_{xx}$ reaches a minimum value, and then increases sharply with the further decrease of temperature. Below 20K, $\rho_{xx}$ starts to saturate and the residual resistivity increases drastically from 0.082 mΩcm to 0.278 mΩcm. When $H$ increases above 1T, this compound begins to demonstrate the semiconducting-like resistivity behavior over the whole temperature range from 300K to 2K. This phenomenon is reminiscent of the metal-semiconductor transition induced by external magnetic field, which is mostly observed in zero-gap materials, such as $WTe_2$,[4] $NbSb_2$,[29] $Cd_3As_2$,[1] and $Bi_2Se_3$.[18] That may be due to their low energy for electron transition from valence band to conductance band which makes resistance quite sensitive to external excitations such as electric or magnetic field, and pressure. When a metal-semiconductor transition appears, the external magnetic field has opened a band gap between valance band and conduction band,[18, 29] which can lead to a drastic increase of resistance and therefore a large positive magnetoresistance MR (MR = $[\rho(H)-\rho(0)]/\rho(0)]\times 100\%$). In Figure 2 (b), the corresponding value of MR deduced from Figure 2 (a) was plotted as a function of temperature. At 2K, this compound exhibits a large MR of 320% in 10T. With the increase of temperature, MR is suppressed significantly due to the thermal fluctuation but still kept around 102% at 300K. More intriguingly, we have observed that MR increases sharply in the low field region below 100K, and a large MR of 240% is obtained in a magnetic field of 1T at 2K. Notably, the MR in ScPtBi single crystal at low-field region is much larger than that of many other nonmagnetic compounds, such as $Ag_{2\pm\delta}Te$,[30-32] $Bi_2Se_3$,[18, 33] $\beta$-CuAgSe,[34] $PtSn_4$,[35] $LaAgBi_2$,[36] NiMnSb,[37] and LuPdBi,[16] which makes it a potential material candidate for applications in high-sensitivity magnetic sensors.

To offer a further insight of the relationship between external magnetic field and MR, the magnetic field dependence of MR at selected temperatures is shown in Figure 2 (c). It is of interest to notice that MR shows a pronounced cusp in low-field region below 100 K. The extraordinary MR can be associated with the WAL effect which can arise from both the strong spin-orbit coupling (SOC) and surface states in topological insulators.[38] Since ScPtBi has been predicted to be topologically nontrivial semimetal, it was worth investigating whether the observed WAL effect originates from the two dimension (2D) surfaces states. The response of magnetoconductance $\Delta G = G(H) - G(0)$ was fitted in the term of the Hikami-Larkin-Nagaoka (HLN) formula,[39]

$$\Delta G(H) = \frac{\alpha e^2}{2\pi^2 \hbar}[\psi(\frac{1}{2} + \frac{\hbar}{4eHL_\varphi}) - \ln(\frac{\hbar}{4eHL_\psi})] \quad (1)$$

where $\psi$ is the digamma function, $L_\varphi$ is the phase coherence length, and $\alpha$ is the prefactor which is experimentally found to be equal to -0.5 for each conductive channel in a traditional 2D electron system. In Figure 3 (a), we present the low-field $\Delta G$ at temperatures ranging from 2K to 40K. It is found that $\Delta G$ is fitted well using the HLN model and the corresponding values of $L_\varphi$ and $\alpha$ are shown in Figure 3 (b) and the inset, respectively. We can see that the values of $L_\varphi$ decrease correspondingly with the increase of temperature, suggesting the weakening of WAL effect at high temperatures. On the other hand, the values of coefficient $\alpha$ are in the order of $10^6$, which is much larger than that of the 2D system. It should be noted here that the observation of the huge $\alpha$ is not extraordinary and similar results have been obtained in the recently reported LuPdBi,[6, 16] YPtBi,[8] and LuPtSb.[15] This indicates the WAL effect in ScPtBi single crystal mainly originates from the 3D bulk contribution. In order to further confirm the 3D-domainted WAL effect, the angle-dependent $\Delta G$ was measured at 2K in a field limit of 1 T. In Figure 3(c), we show the dependence of $\Delta G$ on the perpendicular component of magnetic field $H \sin\theta$ in different angles. For a 2D-domainted WAL, the $\Delta G$ vs $H \sin\theta$ curves should overlap onto one curve. However, the four curves deviate from each other, further proving that the WAL effect in ScPtBi single crystal is mainly due to the 3D bulk transport but not the 2D surface states. As mentioned above, the field-induced metal-semiconductor-like transition suggest that the external magnetic field opens a gap between the valence and conduct band in our ScPtBi single crystal, however the huge $\alpha$ together with the angle-dependent magnetoconductivity indicate that the 2D surface states are still deeply buried in the 3D bulk ones. In this regard, it is necessary to detect the possible surface states (such as linear MR and 2D Shubnikov-de Hass quantum oscillations) in high magnetic field region for our further work.

We have further carried out the Hall measurement at temperatures ranging from 300K to 2K in a field limit of 8T to determine the mobility. Figure 4 (a) displays the magnetic field dependence of Hall resistivity $\rho_{xy}$ at 300K and 2K. Both of them show positive linear function of the magnetic field up to 8T, demonstrating that only one type of hole carrier dominates the transport properties. Based on the one-band model, the Hall mobility $\mu_h$ and carrier concentration $n$ of ScPtBi single crystal can be established, and their corresponding values are plotted as a function of temperature in Figure 4 (b) and (c), respectively. The deduced carrier concentration falls within the range of $2.1\times10^{19}$ cm$^{-3}$ (300K)-$1.8\times10^{19}$ cm$^{-3}$ (2K), which is in the same order with other rare-earth based half-Heusler compounds and reveals its semimetal nature.[6, 16, 17] As shown in Figure 4 (c), quite a high mobility of 4050 cm$^2$V$^{-1}$s$^{-1}$ is obtained at 2K. With the increase of temperature, the mobility shows a negative temperature variation, reflecting that the lattice scattering scatting dominates the temperature-dependent mobility. Even at room temperature, this compound exhibits a high value of 2016 cm$^2$V$^{-1}$s$^{-1}$. The observation of high mobility over a wide temperature region up to room temperature in ScPtBi single crystal makes it a promising application in low-power devices.

In summary, we have synthesized the nonmagnetic rare-earth based half-Hesuler ScPtBi single crystals by a Bi-flux method and carried out a thorough study on its magneto-transport. Remarkably, a large MR ranging from 240% at 2K to 150% at 100K was observed in a magnetic field of 1T. The large low-field MR effect can be accounted for two reasons: (i) the field-induced metal-semiconductor transition leads to a large MR effect; (ii) the WAL effect makes resistance increase steeply in the low-field region. Moreover, Hall measurement reveals ScPtBi single crystal a semimetal with high mobility over a wide temperature region even up to room temperature (4050 cm$^2$V$^{-1}$s$^{-1}$ at 2K - 2016 cm$^2$V$^{-1}$s$^{-1}$ at 300K). That may be due to its nearly linear dispersion of band structure where the charge carriers behave like massless particles. The combination of large low-field MR with the high mobility in ScPtBi single crystal makes it a potential material candidate for applications in low-power and high-sensitivity magnetic sensors.

**Acknowledge**

This work is supported by the National Basic Research Program of China (973 Program 2012CB619405), National Natural Science Foundation of China (Grant Nos. 11474343 and 11574374), and Strategic Priority Research Program B of the Chinese Academy of Sciences under the grant No. XDB07010300.

**References**

[1] T. Liang, Q. Gibson, M. N. Ali, M. H. Liu, R. J. Cava, and N. P. Ong, Nat. Mater. **14**, 280 (2014).

[2] X. C. Huang, L. X. Zhao, Y. J. Long, P. P. Wang, D. Chen, Z. H. Yang, H. Liang, M. Q. Xue, H. M. Weng, Z. Fang, X. Dai, G. F. Chen, Phys. Rev. X **5**, 031023 (2015).

[3] C. Shekhar, A. K. Nayak, Y. Sun, M. Schmidt, M. Nicklas, I. Leermakers, U. Zeitler, Y. Skourski, J. Wosnitza, Z. K. Liu, Y. L. Chen, W. Schnelle, H. Borrmann, Y. Grin, C. Felser, B. H. Yan, Nat. Phys. **11**, 645 (2015).

[4] M. N. Ali, J. Xiong, S. Flynn, J. Tao, Q. D. Gibson, L. M. Schoop, T. Liang, N. Haldolaarachchige, M. Hirschberger, N. P. Ong, and R. J. Cava, Nature **514**, 205 (2014).

[5] P. C. Canfield, J. D. Thompson, W. P. Beyermann, A. Lacerda, M. F. Hundley, E. Peterson, Z. Fisk, and H. R. Ott, J. Appl. Phys. **70**, 5800 (1991).

[6] G. Z. Xu, W. H. Wang, X. M. Zhang, Y. Du, E. K. Liu, S. G. Wang, G. H. Wu, Z. Y. Liu, and X. X. Zhang, Sci. Rep. **4**, 5709 (2014).

[7] F. F. Tafti, T. Fujii, A. Juneau-Fecteau, S. René de Cotret, N. Doiron-Leyraud, A. Asamitsu, and L. Taillefer, Phys. Rev. B **87**, 184504 (2013).

[8] N. P. Butch, P. Syers, K. Kirshenbaum, A. P. Hope, and J. Paglione, Phys. Rev. B **84**, 220504(R) (2011).

[9] S. Chadov, X. L. Qi, J. Kübler, G. H. Fecher, C. Felser, and S.-C. Zhang, Nat. Mater. **9**, 541 (2010).

[10] H. Lin, L. A. Wray, Y. Xia, S. Xu, S. Jia, R. J. Cava, A. Bansil, and M. Z. Hasan, Nat. Mater. **9**, 546 (2010).

[11] D. Xiao, Y. Yao, W. Feng, J. Wen, W. Zhu, X.-Q. Chen, G. M. Stocks, and Z. Zhang, Phys. Rev. Lett. **105**, 096404 (2010).

[12] X. M. Zhang, W. H. Wang, E. K. Liu, G. D. Liu, Z. Y. Liu, and G. H. Wu, Appl. Phys. Lett. **99**, 071901 (2011).

[13] W. Al-Sawai, H. Lin, R. S. Markiewicz, L. A. Wray, Y. Xia, S.-Y. Xu, M. Z. Hasan, and A. Bansil, Phys. Rev. B **82**, 125208 (2010).

[14] W. H. Wang, Y. Du, G. Z. Xu, X. M. Zhang, E. K. Liu, Z. Y. Liu, Y. G. Shi, J. L. Chen, G. H. Wu, and X. X. Zhang, Sci. Rep. **3,** 2181 (2013)**.**

[15] Z. P. Hou, Y. Wang, G. Z. Xu, X. M. Zhang, E. K. Liu, X. K. Xi, W. Q. Wang, W. H. Wang, G. H.

Wu, Appl. Phys. Lett. **106**, 102102 (2015).

[16] O. Pavlosiuk, D. Kaczorowski, and P. Wiśniewski, Sci. Rep. **5**, 9158 (2015).

[17] C. Shekhar, S. Ouardi, A. K. Nayak, G. H. Fecher, W. Schnelle, and C. Felser, Phys. Rev. B **86**, 155314 (2012).

[18] X. L. Wang, Y. Du, S. X. Dou, and C. Zhang, Phys. Rev. Lett. **108**, 266806 (2012).

[19] R. Shan, S. Ouardi, G. H. Fecher, L. Gao, A. Kellock, K. P. Roche, M. G. Samant, C. E. ViolBarbosa, E. Ikenaga, C. Felser, and S. S. P. Parkin, Appl. Phys. Lett. **102**, 172401 (2013).

[20] C. Shekhar, S. Ouardi, G. H. Fecher, A. K. Nayak, C. Felser, and E. Ikenaga, Appl. Phys. Lett. **100**, 252109 (2012).

[21] S. J. Patel, J. K. Kawasaki, J. Logan, B. D. Schultz, J. Adell, B. Thiagarajan, A. Mikkelsen, and C. J. Palmstrøm, Appl. Phys. Lett **104**, 201603 (2014).

[22] P. Villars & L. D. Calvert, Person's Handbook of Crystallographic data for intermetallic phase. (ASM International, 1991).

[23] R. Gautier, X. W. Zhang, L. H. Hu, L. P. Yu, Y. Y. Lin, T. O. L. Sunde, D. Chon, K. R. Poeppelmeier, and A. Zunger, arXiv: 1412.2398.

[24] See supplementary material at [URL will be inserted by AIP] for the EDX spectrum.

[25] E. D. Mun, S. L. Bud'ko, C. Martin, H. Kim, M. A. Tanatar, J.-H. Park, T. Murphy, G. M. Schmiedeshoff, N. Dilley, R. Prozorov, and P. C. Canfield, Phys. Rev. B **87**, 075120 (2013).

[26] R. Shan, S. Ouardi, G. H. Fecher, L. Gao, A. Kellock, A. Gloskovskii, C. E. ViolBarbosa, E. Ikenaga, C. Felser, and S. S. P. Parkin, Appl. Phys. Lett. **101**, 212102 (2012).

[27] S. Ouardi, G. H. Fecher, C. Felser, J. Hamrle, K. Postava, and J. Pištora, Appl. Phys. Lett. **99**, 211904 (2011).

[28] M. L. Tian, J. Wang, W. Ning, T. E. Mallouk, and M. H. W. Chan, Nano. Lett. **15**, 1487 (2015).

[29] K. F. Wang, D. Graf, L. J. Li, L. M. Wang, and C. Petrovic, Sci. Rep **4**, 7328 (2014).

[30] R. Xu, A. Husmann, T. F. Rosenbaum, M. -L. Saboungi, J. E. Enderby, and P. B. Littlewood, Nature **390**, 57 (1997).

[31] H. S. Schnyders, Appl. Phys. Lett **107**, 042103 (2015).

[32] Y. Sun, M. B. Salamon, M. Lee, and T. F. Rosenbaum, Appl. Phys. Lett **82**, 1440 (2003).

[33] Y. Yan, L. X. Wang, D. P. Yu, and Z. M. Liao, Appl. Phys. Lett **103**, 033106 (2013).

[34] S. Ishiwata, Y. Shiomi, J. S. Lee, M. S. Bahramy, T. Suzuki, M. Uchida, R. Arita, Y. Taguchi, and Y. Tokura, Nat. Mater. **12**, 512 (2013).

[35] E. Mun, H. Ko, G. J. Miller, G. D. Samolyuk, S. L. Bud'ko, and P. C. Canfield, Phys. Rev. B **85**, 035135 (2012).

[36] K. F. Wang, D. Graf, and C. Petrovic, Phys. Rev. B **87**, 235101 (2013).

[37] W. R. Branford, S. K. Clowes, M. H. Syed, Y. V. Bugoslavsky, S. Gardelis, J. Androulakis, J. Giapintzakis, C. E. A. Grigorescu, A. V. Berenov, S. B. Roy, and L. F. Cohen, Appl. Phys. Lett. **84**, 2358 (2004).

[38] B. Xia, P. Ren, A. Sulaev, P. Liu, S.-Q. Shen, and L. Wang, Phys. Rev. B **87**, 085442 (2013).

[39] S. Hikami, A. I. Larkin, Y. Nagaoko, Prog. Theor. Phys. **63**, 707 (1980).

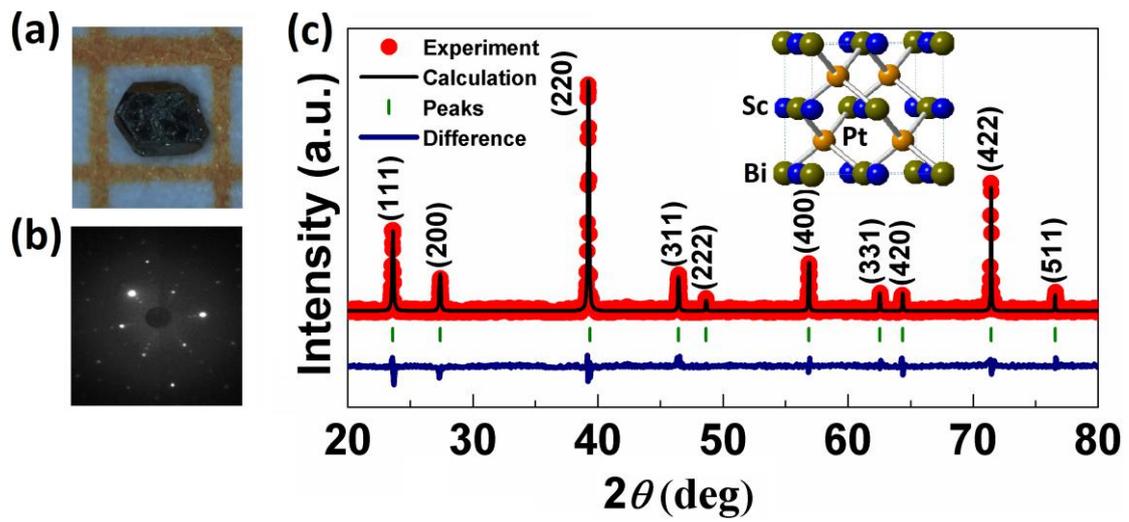

**Figure. 1** (a) The typical photograph of ScPtBi single crystal placed on a millimeter grid. (b) Laue pattern with the incident axis coincident with the [111] zone axis. (c) XRD patterns of single crystals (red circle) and structural refinement results (black line). The differences between observed and calculated profiles are presented by the blue trace. The green segments show the expected diffraction peaks. The inset shows a structure view of conventional ScPtBi unit cell which has 4 number of formulary units.

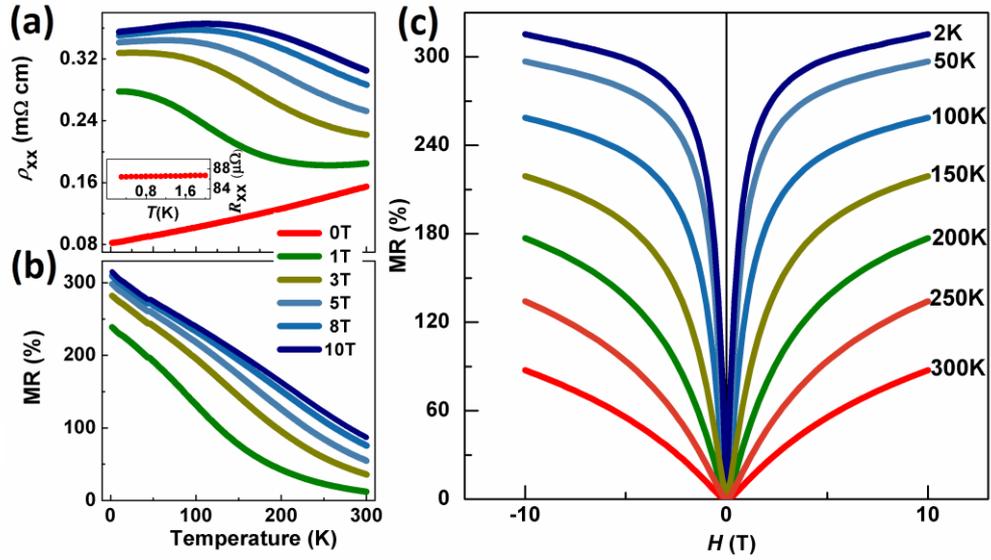

**Figure. 2** (a) Temperature dependence of longitudinal resistivity $\rho_{xx}$ at temperatures ranging from 300K to 2K under different external magnetic fields. Insert: low temperature-dependent $\rho_{xx}$ in zero field reveals ScPtBi does not become superconducting even when the temperature decreases to 0.3K. (b) MR at different magnetic field plotted as a function of temperature. (c) The magnetic field dependence of MR at selected temperatures.

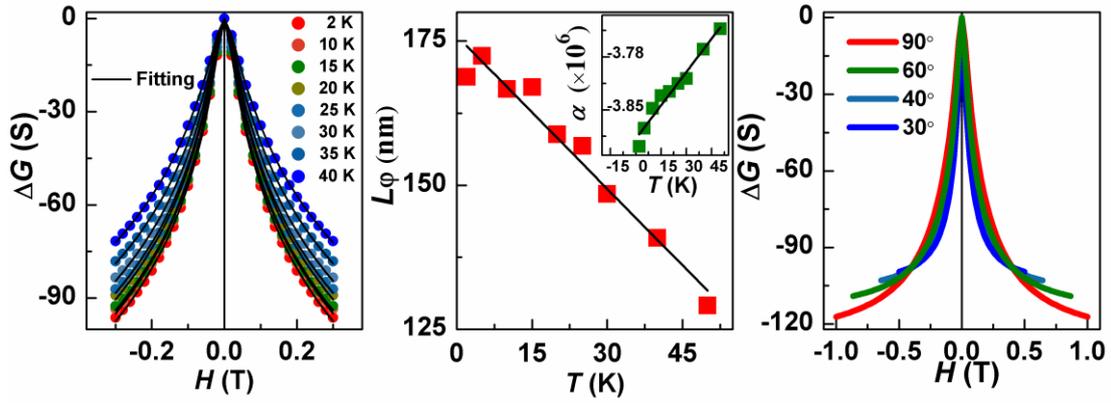

**Figure. 3** (a) Δ*G* curves at a series of temperatures (color circles). The HNL fitting lines are presented in the black solid lines. (b) The fitted values of $L_\varphi$ at the temperature regime 2K – 40K. Inset: the dependence of $\alpha$ on the temperature. (c) The curves of magnetoconductivity vs the perpendicular component of magnetic field. Inset: the schematic of measurement where $\theta$ suggests the angle between the magnetic field and the (111) plane. $\theta = 90°$ means the magnetic field is parallel to (111) plane.

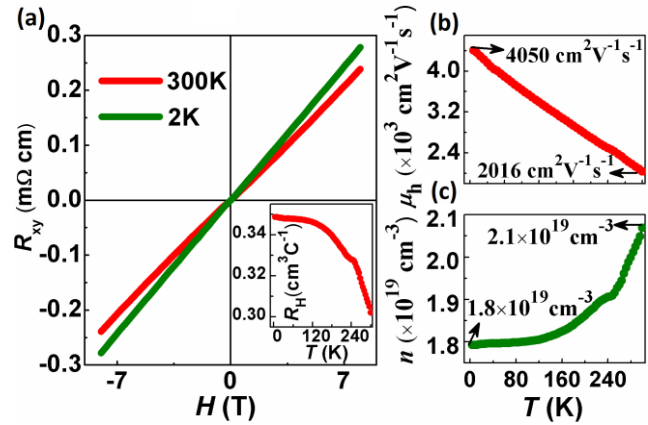

**Figure. 4** (a) Magnetic field dependence of Hall resistivity $\rho_{xy}$ at 300K and 2K in a field limit of 8T. Inset: Temperature dependence of Hall coefficients $R_H$ measured in 8T. (b) Temperature dependence of mobility $\mu_h$. (c) Temperature variation of carrier concentration *n*.